\documentclass{article}

\def\be{\begin{equation}}       \def\eq{\begin{equation}}
\def\ee{\end{equation}}         \def\eqe{\end{equation}}

\def\bea{\begin{eqnarray}}      \def\eqa{\begin{eqnarray}}
\def\ena{\end{eqnarray}}        \def\eea{\end{eqnarray}}
                                \def\eqae{\end{eqnarray}}

\def\ba{\begin{array}}
\def\ea{\end{array}}
\def\unit{1 \hskip-.3em \raise2pt\hbox{$ \scriptstyle |$ } }
\def\a{\alpha}
\def\b{\beta}
\def\c{\gamma} 
\def\d{\delta}
\def\e{\epsilon}           % Also, \varepsilon
\def\f{\phi}               %      \varphi
\def\g{\gamma}

\def\j{\psi}
\def\m{\mu}

\def\r{\rho}                                     %     \varrho
\def\s{\sigma}                                   %     \varsigma

\def\G{\Gamma}

\def\O{\Omega}

\def\cq{{\cal Q}}

\def\half{{1 \over 2}}

\def\pa{\partial}                              % curly d
\def\>{\rangle} %right angle

\def\<{\langle} %left angle
\def\Dsl{D \hskip-.6em \raise1pt\hbox{$ / $ } }

\def\PRD{Phys. Rev. D}
\def\PRL#1#2#3{{\it Phys. Rev. Lett.} {\bf#1} (#2) #3}

\def\NPB#1#2#3{{\it Nucl. Phys.} {\bf B#1} (#2) #3}

\def\PRD#1#2#3{{\it Phys. Rev.} {\bf D#1} (#2) #3}

\def\PLB#1#2#3{{\it Phys. Lett.} {\bf #1B} (#2) #3}

\def\1ov4{{1\over 4}}

\def\pa{\partial}
\def\xx{\times}

\def\pa{\partial}

\def\xx{\times}

\def\nonu{\nonumber \\{}}
\def\half{{1 \over 2}}

\begin{document}

\thispagestyle{empty}
\begin{flushright}
{\sc KUL-TF}-97-17\\
hep-th/9706192
\end{flushright}
\vspace{1cm}
\setcounter{footnote}{0}
\begin{center}
{\LARGE\sc{ Duality and asymptotic geometries
    }}\\[14mm]

\sc{Harm Jan Boonstra\footnote{e-mail: harm.boonstra@fys.kuleuven.ac.be},
    Bas Peeters\footnote{e-mail: bas.peeters@fys.kuleuven.ac.be} and
    Kostas Skenderis\footnote{e-mail:
kostas.skenderis@fys.kuleuven.ac.be},}\\[5mm] 
{\it Instituut voor Theoretische Fysica\\
KU Leuven\\
Celestijnenlaan 200D\\
3001 Heverlee, Belgium}\\[20mm]

{\sc Abstract}\\[2mm]

\end{center}

\noindent 

We consider a series of duality transformations that leads to a
constant shift in the harmonic functions appearing in the description
of a configuration of branes. This way, for several intersections of
branes, we can relate the original brane configuration which is
asymptotically flat to a geometry of the type $adS_k \xx E^l \xx S^m$.
The implications of our results for supersymmetry enhancement,
M(atrix) theory at finite N, and for supergravity theories in diverse
dimensions are discussed.

\vfill

\newpage

Duality symmetries are one of the cornerstones of the second
superstring revolution\cite{wit}.  Using a web of duality symmetries
one can connect all known string theories and eleven dimensional
supergravity, leading to a conjectural master theory that contains all
others as special limits (for reviews see \cite{rev}).  A candidate
for this master theory, the M(atrix) theory, has been recently
proposed in \cite{matrix}.  Geometry and/or topology change of
spacetime due to dualities is an extensively studied subject.  A less
well studied facet of this topic is the change of asymptotic geometry
due to dualities. Some examples of this type have been studied before
in \cite{topch,bakas}.

We shall show in this article using a set of duality transformations
and coordinate transformations that one can map supersymmetric
configurations of $M$-branes and $D$-branes that are asymptotically
flat to configurations that are of the type $adS_k \xx E^l \xx S^m$,
where $adS_k$ is the $k$-dimensional anti-de Sitter space, $E^l$ is
the $l$-dimensional Euclidean space and $S^m$ is the the
$m$-dimensional sphere. As we shall see, one can use a specific
combination of duality and coordinate transformations that we shall
call the shift transformation in order to change the constant part of
the harmonic functions appearing in the $D$ and $M$-brane solutions.

These results have a variety of applications. First of all they lead
to a better understanding of the issue of supersymmetry enhancement
near the horizon of certain black hole configurations.  In addition,
using our results, we shall be able to relate supergravity
computations in the standard space-like compactification of 
$M$-theory to finite N M(atrix) theory.
Also, our results imply that certain spontaneous
compactifications of 10 and 11 dimensional supergravity should
exist. In many cases, one can check that such compactifications indeed
appear in the supergravity literature.

To illustrate the mechanism that converts an asymptotically flat space
to an anti-de Sitter space we will start by considering the
Dabholkar-Harvey solution of $N{=}1$ $10d$ supergravity\cite{dabh}
that describes a fundamental string.  We will then generalize this
result to a non-extremal string and afterwards to an arbitrary
supersymmetric configuration of orthogonally intersecting branes with
at least three transverse directions.

The metric, antisymmetric tensor and dilaton field of the solution that 
describes a fundamental string are given by the following expressions
\bea
&&ds^2 = H(r)^{-1} (-dt^2 + dx_1^2) + (dx_2^2 + \cdots + dx_9^2), \nonu
&&B_{01} = H(r)^{-1} -1; \ \ e^{-2\f}=H(r)  \nonu
&&H(r)= 1 + \frac{\cq}{r^6}; \ r^2=x^2_2 + \cdots + x^2_9, \label{fund}
\eea
H is a harmonic function that depends only on the transverse 
directions.

Let us perform a $T$-duality in the $x_1$ direction. 
Using Buscher's rules one obtains
\bea
&&ds^2 = (H-2) dt^2 + H dx_1^2 + 2 (1-H) dt dx_1 + (dx_2^2 + \cdots +
dx_9^2), \nonu
&& B=0; \ \ e^{-2\f}=1
\eea
We now make a change of coordinates that amounts to
an $SL(2,R)$ transformation
\be
\left(
\begin{array}{c}
t \\
x_1
\end{array}
\right) = 
\left(
\begin{array}{cc}
\a & \b  \\
\g & \d
\end{array}
\right)
\left(
\begin{array}{c}
t' \\
x_1'
\end{array}
\right); \ \ \a \d - \b \g=1, \ \ \b \neq \d. \label{sl}
\ee
Finally, we perform another $T$-duality transformation 
in the $x_1'$ direction\footnote{$T$-duality along a linear
combination of space-like coordinates has been considered in
\cite{BMM}.}. The result is (\ref{fund}), but with  
the harmonic function replaced by
\be
H \rightarrow (\b -\d) ((\b -\d) H - 2 \b) \label{res1}
\ee
and, in addition, the constant part of $B_{01}$ is now equal to 
$(\a - \g)/(\b -\d)$, which can be 
shifted to $1$ by a constant gauge transformation (for generic
$SL(2,R)$ transformations there may be global obstructions to these
gauge transformations; these are absent for the cases considered below).
It follows that if one chooses the $SL(2,R)$ with $\d=-\b$ 
the resulting configuration has harmonic function with zero constant 
part and therefore different asymptotics. If in addition we demand that 
the charge $\cq$ remains unchanged then this fixes the parameters (up to
an overall sign) to be $\d=-\b=1/2$. This still leaves a one-parameter family
of $SL(2,R)$ transformations, given by  $\a + \g=2$, that yields the same
result. There is, however, a global issue that we have not yet discussed.
In order to perform the $T$-duality in the $x_1$ direction we 
need to compactify this direction. Assuming that $t$ is non-compact,
$(t, x_1)$ have the geometry of a cylinder. Generically, an 
$SL(2,R)$ transformation is not well-defined on a cylinder. However,
there is a member in the one-parameter family of $SL(2,R)$ transformations 
given above that is well-defined. This is the case for $\g=0, \a=2$.
In this case, 
\be
t'={1 \over 2} (t + x_1) \ \ \ x_1' = 2 x_1.
\label{pref}
\ee
Notice that both $T$-dualities that are involved in the shift transformations
are along a space-like direction.  
A similar set of transformations (with $\g=0$) 
has been used by Hyun\cite{hyun} in order to connect $4d$ and $5d$ 
black holes to the three-dimensional BTZ black hole\cite{BTZ}. 
We also note that the above family 
of $SL(2,R)$ transformations contains the case, 
$\a=\g=1$, which brings one from $(t, x_1)$ to the light-cone frame. 
Finally, notice that the transformations just described can be thought of as 
part of an $O(2,2)$ $T$-duality group associated with the isometries in
the $(t,x_1)$ directions.

Let us generalize this result to non-extremal solutions.
A simple algorithm that leads to a non-extreme version of a given 
extreme solution has been given in \cite{ts1}. The mechanism 
that we just described for the extremal case generalizes straightforwardly
to the non-extremal case. 
To illustrate the method we shall discuss the non-extreme string 
solution of \cite{HorStrom} (which can also be viewed as 
a double dimensional reduction of the non-extremal
$11d$ membrane solution of G\"{u}ven\cite{guven}).
The metric, antisymmetric tensor and dilaton field are given by
\bea 
&&ds^2 = H(r)^{-1}(-f(r)dt^2+dx_1^2) + f(r)^{-1} dr^2 + r^2 d \O_7^2, \nonu
&&B_{01}= H'(r)^{-1} -1, \ \ e^{-2\f} = H(r), \ \ r^2 = x_2^2 + \cdots
+ x_9^2
\label{fund1}
\eea 
where 
\bea
&& H'(r)^{-1}= 1 - \frac{Q}{r^6} H^{-1}; \ \ Q=\m \sinh \psi \cosh \psi; \nonu
&& H(r)= 1 + \frac{\cq}{r^6}; \ \ \cq=\m \sinh^2 \psi; \nonu
&& f(r)=1 - \frac{\m}{r^6}
\eea
The extreme solution is found in the limit $\m \rightarrow 0$, 
$\psi \rightarrow \infty$, while the charge $Q$ is kept fixed.
In this limit one gets $H'=H, \cq=Q, f=1$, and the solution (\ref{fund1})
collapses to the extreme one in (\ref{fund}).

One can now perform the set of transformations $T_1 SL(2,R) T_1$ 
(we use the notation $T_i$ to denote the $T$-duality transformation in the
$x_i$ direction) exactly in the same way as for the extremal case.  
The result is that one can remove the constant part of the harmonic
function $H$ provided that $\d=-\b=(1+ \coth \psi)^{-1}$, where the 
last equality is required only if one wants to preserve the values of the
charges. In addition, $B_{01}$ is transformed (again up to 
constant gauge transformation) to $B_{01} \rightarrow (H-1)^{-1} -1$.
The parameters $\a$ and $\g$ are still restricted  
by the determinant condition, $\a \d -\b \g =1$. 
A similar set of transformations (with $\g=0$) has been considered previously 
in \cite{hyun}.
One easily checks that the values of $\b, \d$ coincide
with the extremal ones in the extremal limit.

We now show that a configuration consisting of four or
fewer fundamental strings, solitonic fivebranes, $D$-branes,
waves and Kaluza-Klein monopoles in type II theory, subject to the
conditions below, can be mapped by
means of a series of duality transformations to a specific
configuration of four (or fewer) $D3$-branes. It is then shown that
the harmonic functions in the resulting configuration can be shifted
in a similar fashion as shown above. We assume throughout
that all harmonic functions depend only on overall transverse
directions and that there are at least three of the latter,
so that the harmonic functions are bounded at
infinity. Furthermore, we require the
configurations to be built according to the intersection rules
based on the `no-force' condition (\cite{Tse1} and
references therein). 
Together, these requirements imply that there
are at most four independent charges. Moreover, the
fraction of supersymmetry preserved is $1/2^n$ for $n\leq3$
and $1/8$ for $n=4$ if $n$ is the number of charges.

Starting with an arbitrary set of branes, waves and
monopoles in type 
IIA theory, subject to the above conditions, the allowed
configurations contain at most one wave and one monopole\cite{Ber}.
By making a $T$-duality transformation these can be mapped into a
fundamental string and a solitonic fivebrane, respectively. The
configuration with four charges (the case with fewer charges can be
dealt with similarly) can now be related to four $D3$-branes
in type IIB: each 
brane can be turned into a $D$-brane using $S$-duality, and can then
be mapped into a $D3$-brane using $T$-duality.
At each step in this procedure there is enough freedom to
choose the directions along which to $T$-dualize to make sure that the
$D3$-branes already present remain $D3$-branes (which are then inert
under $S$-duality), and to avoid mapping any of the other branes into
a wave or a monopole.

In order to show that the harmonic functions in the resulting
configuration of four $D3$-branes can
each be shifted by a constant term, let us choose coordinates such
that the branes are in the $(x_1,x_2,x_3)$, $(x_1,x_4,x_5)$,
$(x_2,x_4,x_6)$ and $(x_3,x_5,x_6)$ directions, respectively; there are
three overall transverse coordinates $x_7,x_8,x_9$. We now make a
$T_{23}$ and $S$ transformation to obtain a fundamental string, a
solitonic fivebrane, and two $D3$-branes described by the following
background fields
\bea
ds^2 &=& H_1^{-1} H_3^{-\half} H_4^{-\half} \left( -dt^2 \right) +
H_1^{-1} H_3^{\half} H_4^{\half} \left( dx_1^2 \right) \nonu
&& \hspace{-1cm} {}+ H_3^{\half} H_4^{-\half} \left( dx_2^2 + dx_5^2 \right) 
+ H_3^{-\half}
H_4^{\half} \left( dx_3^2 + dx_4^2 \right) + H_2^{} H_3^{-\half} H_4^{-\half}
\left( dx_6^2 \right) \nonu
&& \hspace{-1cm} {}+ H_2 H_3^{\half} H_4^{\half} \left( dx_7^2 + dx_8^2 
+ dx_9^2
\right) \nonu
&& \hspace{-1.5cm} B_{01} = H_1^{-1} -1 \qquad H_{ijk} = (dB)_{ijk} 
= \e_{ijkl} \pa_l
H_2 \qquad (i,j,k,l = 6,7,8,9) \nonu
&& \hspace{-1.5cm} A_{0346} = - H_3^{-1}  \qquad A_{0256} = - H_4^{-1} \nonu
&& \hspace{-1.5cm} e^{-2\f} = H_1 H_2^{-1} 
\eea
The $H_i$ are harmonic functions, depending only on the overall
transverse directions.
With this configuration we can again perform the same steps as before,
i.e. make a $T$-duality transformation in the $x_1$ direction,
followed by a change of coordinates, and then again a $T$-duality
transformation in the $x_1$ direction. Because this direction is
along the solitonic fivebrane, and orthogonal to both $D$-branes, the
corresponding harmonic functions appear with the appropriate powers
for this procedure to result in just changing the harmonic function
$H_1$ by a constant shift as before. We can now return to the initial
configuration by taking the $T_{23}S$-dual. As this configuration is
completely symmetric in the four $D3$-branes, we can similarly shift
any other harmonic function by a constant.

The shift transformations in the harmonic functions are particularly
interesting for the case of brane configurations with some simple
near horizon geometry. Such branes interpolate between this geometry
and Minkowski spacetime at infinity.
In eleven dimensions, the membrane ($M2$) interpolates
between ${\cal M}_{11}$ and $adS_4\times S^7$, whereas the
fivebrane ($M5$) interpolates between ${\cal M}_{11}$ and $adS_7\times
S^4$\cite{GiTo}.
We may dimensionally reduce the $M2$-brane to a fundamental (type IIA)
string,  do the shift procedure as explained before, and lift this
configuration up to eleven dimensions again. This will give the
$adS_4\times S^7$ solution of eleven-dimensional
supergravity\footnote{This corresponds to the Freund-Rubin
compactification\cite{FR} to four dimensions with maximal supersymmetry.},
which is then of course just the $M2$-brane solution with the harmonic
function not containing the constant term. For the $M5$-brane one can
similarly show that it is connected via dualities and a simple coordinate
transformation to the $adS_7\times S^4$ geometry.
In fact, $adS_4\times S^7$ and $adS_7\times S^4$ are known to be
maximally supersymmetric vacua of $d{=}11$ supergravity.
There is one ten-dimensional $p$-brane with similar properties:
the self-dual threebrane. Its near horizon geometry, $adS_5\times S^5$,
is itself a maximally supersymmetric vacuum of type IIB supergravity.
All this raises the question, however, how two such solutions can be
related by symmetry transformations, as one of them (the brane)
breaks half the supersymmetry and the other is maximally
supersymmetric. We get back to this paradox after we have discussed
supersymmetry enhancement.

Let us consider orthogonal $M$-brane intersections with
a simple near horizon geometry which is a product containing an
anti-de Sitter spacetime and a sphere. There is only one intersection
of two $M$-branes 
which falls into this class, namely the $M2\perp M5$ solution,
\bea
&&ds^2=H_2^{-\frac{2}{3}}H_5^{-\frac{1}{3}}(-dt^2+dx_1^2)
    +H_2^{-\frac{2}{3}}H_5^{\frac{2}{3}}(dx_2^2)
    +H_2^{\frac{1}{3}}H_5^{-\frac{1}{3}}(dx_3^2+\cdots +dx_6^2)\nonu
    &&\ \ \ \ \ \ 
    +H_2^{\frac{1}{3}}H_5^{\frac{2}{3}}(dx_7^2+\cdots +dx_{10}^2)\,,
\label{25metric}\\
&&F_{r012}= \pm \partial_r H_2^{-1}\ \,,\ \ \ \ \ 
F_{2\a\b\c}=
% \pm \bar{\e}_{\a\b\c}\partial_r H_5=
% -2\pm Q_2^{-1/2}\e_{\a\b\c}\,,
 \pm \e_{\a\b\c}\partial_r H_5\,,\nonumber
\eea
where the signs differentiate between brane and anti-brane,
and $\e_{\a\b\g}$  is the volume form of the three-sphere surrounding
the intersection.
The harmonic functions are
\be\label{harm}
H_2=1+ {Q_2\over r^2}\ ,\ \ \ \ 
H_5=1+ {Q_5\over r^2}\ ,
\ee
where $r^2=x_7^2+x_8^2+x_9^2+x_{10}^2$. With this choice of
harmonic functions the solution is asymptotically Minkowski.
The near horizon geometry is the geometry for $r\rightarrow 0$. In
this limit the constant parts of the harmonic functions become negligible.
Of course, (\ref{25metric}) with harmonic functions without the
constant term is still a solution of the supergravity.
The corresponding geometry can
also be obtained throughout spacetime by performing the shift
transformation. We get 
\bea
ds^2 &=& Q_2^{-\frac{2}{3}}Q_5^{-\frac{1}{3}}r^2(-dt^2+dx_1^2)
    +Q_2^{-\frac{2}{3}}Q_5^{\frac{2}{3}}(dx_2^2)
    +Q_2^{\frac{1}{3}}Q_5^{-\frac{1}{3}}(dx_3^2+\cdots +dx_6^2)\nonu
&&\ \ \ \ \ \ 
     +Q_2^{\frac{1}{3}}Q_5^{\frac{2}{3}}({1\over r^2}dr^2 +d\Omega_3^2)\,.
\eea
Thus the spacetime factorizes into the product of an
$adS_3$ spacetime (with coordinates $t,x_1,r$), a three-sphere $S^3$
of radius $Q_2^{\frac{1}{6}}Q_5^{\frac{1}{3}}$, and a flat Euclidean
five-dimensional space $E^5$.
Similarly, one finds the other possible orthogonal
intersections that give rise to this kind of geometry: for three
charges these are the $M2\perp M2\perp M2$ and
$M5\perp M5\perp M5$ (with three overall transverse directions),
and for four charges the $M2\perp M2\perp M5\perp M5$ intersection.
Table 1 shows the corresponding geometries.
These near horizon geometries have also been given in \cite{KK}.
Being solutions to the field equations, the geometries in table 1
can also be interpreted as spontaneous
compactifications of eleven-dimensional supergravity on
a sphere and/or a torus (after periodically identifying Euclidean
coordinates).

\begin{center}
\vspace{.2cm}
\begin{tabular}{|c|c|}
\hline
$M2\perp M5$ & $adS_3\times E^5\times S^3$\\ \hline
$M2\perp M2\perp M2$ & $adS_2\times E^6\times S^3$\\ \hline
$M5\perp M5\perp M5$ & $adS_3\times E^6\times S^2$\\ \hline
$M2\perp M2\perp M5\perp M5$ & $adS_2\times E^7 \times S^2$\\
\hline
\end{tabular}

\vspace{.2cm}
{\bf Table 1}
\end{center}
\vspace{.1cm}

The product geometries have the form
$adS_{p+2}\times E^{q}\times S^{9-p-q}$, where $p$ is the spatial
dimension of the intersection, $q$ is the number of relative transverse
coordinates and $9-p-q$ is the number of overall transverse coordinates
minus one.
A wave can be added to the common string of the $M2\perp M5$
and $M5\perp M5\perp M5$ intersections (as well as
to the single $M2$ and $M5$-branes).
It modifies only the $adS_3$ part of the corresponding geometry.
After reduction to ten dimensions, the $M2\perp M5$ intersection with
a wave is dual to the $M2\perp M2\perp M2$ intersection,
and the $M5\perp M5\perp M5$ intersection with a wave is dual
to the $M2\perp M2\perp M5\perp M5$ intersection.

It is well-known\cite{Gibb} that some solutions exhibit supersymmetry
enhancement at the horizon. For example, the $M2$ and $M5$-branes
break one half of supersymmetry, whereas their near horizon geometries
$adS_4\times S^7$ and $adS_7\times S^4$ are maximally
supersymmetric vacua of $d=11$ supergravity.
A number of other cases of supersymmetry enhancement for $p$-branes
in different dimensions is known,
and in all these cases the near horizon geometry contains a factor
$adS_{k}\times S^{m}$.
The self-dual threebrane in ten dimensions is
another example.
We find that in fact all intersections of table 1
exhibit supersymmetry enhancement at the horizon.
This becomes clear by the observation that the $adS_{p+2}\times E^q
\times S^{9-p-q}$ solutions themselves preserve twice as much
supersymmetry as the corresponding $M$-brane intersections.
We illustrate this for the $adS_3\times E^5\times S^3$
solution corresponding to the $M2\perp M5$ intersection.

Unbroken supersymmetries correspond to vanishing supersymmetry variations
of the gravitino,
\be\label{susy}
\d\j_M=D_M\varepsilon +{1\over288}\left(\G_M{}^{NPQR}-8\d_M{}^N \G^{PQR}
\right)F_{NPQR}\varepsilon =0\,,
\ee
in the background (\ref{25metric}), with $H_2={1\over r^2}$
and $H_5={1\over r^2}$\footnote{Without loss of generality, we take
$Q_2=Q_5=1$.}.
We split this equation in the three parts corresponding to
$adS_3$ (with coordinates $x^\m$), $E^5$ (coordinates $y^s$)
and $S^3$ (coordinates $z^\a$).
The $\Gamma$-matrices can be conveniently chosen to be
\bea
&&\G_\m=\c_\m\otimes{\bf 1}_{4}\otimes{\bf 1}_{2}
 \otimes\s_1\nonu
&&\G_s={\bf 1}_{2}\otimes\c_s\otimes{\bf 1}_{2}
 \otimes\s_3\\
&&\G_\a={\bf 1}_{2}\otimes{\bf 1}_{4}\otimes\c_\a
 \otimes\s_2\nonumber\,.
\eea
Substituting in (\ref{susy}) one finds for the $E^5$ components:
\be\label{torus}
\d\j_s=\partial_s\varepsilon + {1\over6}{\bf 1}\otimes
 (-\c_s\c^2 +3\d_s{}^2)\otimes{\bf 1}\otimes(\s_1+i\s_2)
 \varepsilon =0\,.
\ee
Writing $\varepsilon=\eta(x)\otimes\xi(y)\otimes\r(z)\otimes\chi$,
we see that
(\ref{torus}) reduces to $\partial_s \xi =0$ if we assume
that the two-component spinor $\chi$ is projected onto its upper
component. Thus, $\xi$ are the constant Killing spinors of $E^5$.
For the $adS_3$ components we get
\be
D_\m\varepsilon -{1\over6}\c_\m\otimes\c^2\otimes{\bf 1}
 \otimes({\bf 1}+2\s_3)\varepsilon =0\,,
\ee
and this yields
\be
D_\m \eta\mp {1\over2}\c_\m\eta=0\ \ {\rm for}\ \ \g^2\xi=\pm\xi\ \ 
{\rm and}\ \ \s_3\chi=\chi\,,
\ee
which coincides with the geometric Killing spinor equation
for $adS_3$.
Finally, the $S^3$ components give
\be
D_\a\varepsilon -{i\over6}{\bf 1}\otimes\c^2\otimes\c_\a
 \otimes({\bf 1}+2\s_3)\varepsilon =0\,.
\ee
This translates into the geometric Killing spinor equation
on $S^3$:
\be
D_\a\r\mp {i\over2}\c_\a\r=0\ \ {\rm for}\ \ \g^2\xi=\pm\xi\ \ 
{\rm and}\ \ \s_3\chi=\chi\,.
\ee
Since $adS_3$, $E^5$ and $S^3$ all admit the maximal number
of Killing spinors and there is only one projection
on $\chi$, we conclude that this solution preserves
one half of supersymmetry which is double the amount
preserved by the $M2\perp M5$ intersection in the bulk.
For the other cases of table 1, we find that (\ref{susy})
in those backgrounds reduces to the appropriate geometric Killing spinor
equations up to two projections, thus establishing supersymmetry
doubling also in these cases\footnote{For the $M2\perp M2\perp M5
\perp M5$ configuration, there is enhancement only if the
four signs related to brane/anti-brane choice multiply to $-1$
(in our conventions); for opposite orientations supersymmetry is
completely broken.}.

All known solutions that exhibit supersymmetry enhancement not
only have a near horizon geometry of the form $adS_{p+2}\times
E^q\times S^{d-p-q-2}$, but also have regular (i.e. finite) dilaton
(if any) near the horizon. This is true for the four and
five-dimensional extremal black holes with nonzero entropy, and
also for the $D3$-brane which doesn't couple to the dilaton.
We observe that, for the eleven-dimensional configurations of table 1,
a dimensional reduction over one or more of the relative
transverse directions will always give rise to a dilaton
in lower dimensions that is regular at the horizon. That is because
the dilaton originates from the eleven-dimensional metric
according to
\be
ds_{11}^2=e^{2\alpha\phi}ds_{10}^2 + e^{2\beta\phi}dy^2\,,
\ee
where $y$ is the coordinate that is reduced over, and $\alpha$
and $\beta$ are constants related to the choice of metric frame and
the normalization of the dilaton.
Since the metric diagonal component for a relative transverse direction
is regular for the configurations of table 1, $e^{2\beta\phi}$
and thus $\phi$ is regular. Reduction over such directions
also does not interfere with supersymmetry. Hence all solutions
that can be obtained in this way must exhibit supersymmetry
enhancement. The same is true for configurations obtained
by further $T$-dualities in the relative transverse directions.

We now turn to the paradox raised by the observation
that the shift transformation relates solutions
with different amounts of unbroken supersymmetry.
For clarity, we take the example of the self-dual threebrane.
It is related via a series of dualities plus a simple coordinate
transformation to the $adS_5\times S^5$ solution of type IIB
supergravity. The latter has maximal supersymmetry, with
Killing spinors that are the direct product of the Killing
spinors of $adS_5$ and $S^5$. However, in order to do the
shift transformation we have to perform $T$-duality along
some coordinates of $adS_5$. To this end we have to
compactify these coordinates. Only half of the Killing
spinors of $adS_5$ spacetime survive compactification as 
one can immediately verify by looking at the explicit expression
of the Killing spinors\cite{LPT}. In other words, $adS_5$ with 
a set of compact isometries (needed in order to perform $T$-dualities)
admits only half of the Killing spinors compared to  $adS_5$
with no compact isometries. Actually, this is true
for all anti-de Sitter spacetimes. A familiar case of this sort 
is that of the BTZ black hole\cite{BTZ}. This three-dimensional black hole
can be obtained by compactifying one of the coordinates of $adS_3$.
The compactification kills all Killing spinors in the non-extremal
case, and half of them in the massless extremal one\cite{CH}. The
rotating extremal case preserves only $1/4$ of the supersymmetry as it
also contains a wave.
A similar situation was also discussed in \cite{LPT}, where it is shown
that a maximally supersymmetric $adS_d$ solution can be reduced to
a half-supersymmetric domain wall solution in $d{-}1$ dimensions.
Let us note here that the case of dualization  
along a rotational isometry is different from the discussion above. 
In spaces with rotational isometries
the Killing spinors also depend on the coordinates along which one dualizes.
Although it appears that the spacetime supersymmetry is lost
after the $T$-duality\cite{bakas,BKO}, it is actually  
non-locally realized\cite{AAB}. In our case, the $T$-duality maps
between two
configurations with the same amount of spacetime supersymmetry. 
Only after decompactification one can detect the enhanced supersymmetry.
This resolves the paradox as the threebrane is a half-supersymmetric solution.
On the other hand, dimensional reduction or $T$-duality along toroidal
directions does not break any supersymmetry. So solutions
obtained by dimensional reduction along relative transverse
directions of the $M$-brane intersections of table 1 should
still exhibit supersymmetry enhancement.

As an example, let us consider the six-dimensional
dyonic string solutions of~\cite{DFKR},
\bea
&&ds^2=H_2^{-1}(-dt^2+dx_1^2)+H_5(dx_7^2+ \cdots +dx_{10}^2)\nonu
&&e^{-2\phi}= H_2 H_5^{-1}\label{ds}\\
&&H_{r01}= \pm\partial_r H_2^{-1}\ \,,\ \ \ \ 
H_{\a\b\c}=\pm \e_{\a\b\c}\partial_r H_5\,,
\nonumber
\eea
written in the six-dimensional string frame.
As suggested by the way of presenting, these dyonic strings are
obtained by double dimensional reductions over the relative transverse
directions of the $M2\perp M5$ intersection.
If electric and magnetic charges are chosen equal ($Q_2=Q_5$ in
(\ref{harm})), (\ref{ds}) becomes the
self-dual string\cite{DL}. The solutions (\ref{ds}) preserve
$1/4$ of supersymmetry, but now we know that this must
be enhanced to $1/2$ at the horizon.
The near horizon geometry of the strings is $adS_3\times S^3$.
The self-dual string can also be embedded into six-dimensional
$N{=}2$ chiral supergravity where it breaks only $1/2$ of supersymmetry,
and hence $adS_3\times S^3$ is a maximally supersymmetric vacuum.

We would now like to discuss the relation between this shift
transformation and results recently obtained in M(atrix)
theory. Following the proposal of \cite{matrix} one needs to
consider $M$-theory in the infinite momentum frame to obtain a
formulation in terms of matrix model quantum mechanics in the large N
limit. Compactifying along a null-direction is conjectured to lead, in the
framework of DLCQ, to a correspondence between $M$-theory and the
Matrix model for finite~N\cite{suss}. In \cite{BBPT} scattering
amplitudes for configurations of $D$-branes and gravitons have been
computed, both in the effective supergravity theories and through the
Matrix model approach.  It was observed that the supergravity
calculation always produces identical results to the Matrix theory in
the infinite N limit, in which subleading corrections are
suppressed. However, only for the supergravity corresponding to
reduction of the $11d$ theory along a light-like coordinate, the
computations also agree for finite N. In this case, the harmonic
functions of the $D0$-branes in the supergravity calculation are
shifted to vanish at 
infinity.  Comparing with our results, we have found a one-parameter
family of `frames' in which this shift occurs, and which therefore also
lead to agreement with the Matrix model computation for finite N.
With `frames' we mean the frames that are obtained after applying the
shift transformations in the Infinite Momentum Frame. This
one-parameter family of frames contains the
$10d$ light-cone frame. However, it is not clear to us whether this is related
to the light-cone frame in the DLCQ.  In all frames considered in
(\ref{sl}), except for (\ref{pref}), the shift transformation can only
be applied in the supergravity limit. 
This is due to global issues we already discussed.
For the transformation
(\ref{pref}) such problems are absent since we do not need to
compactify the time coordinate. Furthermore, the 
$T$-dualities are along space-like directions, and this transformation
should be well-defined also in the full theory. Finally, let us note 
that in order to perform the shift transformation to a $D0$-brane we
need to assume that the background admits at least one spatial
isometry. 

We conclude by making some comments on the implications of our results
for supergravity theories in various dimensions. Each configuration
in $11$ dimensions with effective geometry of the type $adS_k \xx E^l
\xx S^m$ corresponds to a solution of $11d$ supergravity
with the appropriate amount of supersymmetry. It also follows directly
that, after reduction along $p \leq l$ of the flat directions, the
geometry $adS_k \xx E^{l-p} \xx S^m$ is a solution with the same
amount of supersymmetry (counting the number of spinor components) in
$11-p$ dimensions. In addition we can deduce the existence of solutions
with a certain amount of supersymmetry after spontaneous
compactification on the sphere $S^m$. These compactifications are expected
to give
rise to solutions of gauged supergravities in $11{-}m{-}p$ dimensions
with geometry 
$adS_k \xx E^{l-p}$.
Several of these results are well-known, such as
the spontaneous compactification of $11d$ supergravity on $S^7$,
giving gauged $N{=}8$ supergravity in $d{=}4$, and
the $adS_7\times S^4$ and $adS_5\times S^5$ solutions
of $d{=}11$ supergravity and type IIB supergravity. The anti-de Sitter
parts of these solutions are
maximally supersymmetric vacua of gauged maximal supergravities
in seven and five dimensions\cite{gasu}.

Let us also make some remarks about the $10d$ case. As shown
in \cite{GiTo}, the solitonic fivebrane gives rise (in the
string frame) to an asymptotic geometry of the type ${\cal M}_7 \xx
S^3$, whereas the fundamental string yields (in the solitonic
fivebrane metric) an asymptotic geometry of $adS_3 \xx
S^7$\cite{DGT}.  Similarly, we find that for the
$Dp$-branes, in the `dual $Dp$-metric' -- the metric in which the
curvature and the $(8{-}p)$-form field strength appear in the action
with the same power of the dilaton
-- the asymptotic geometry is of the type $adS_{p+2} \xx
S^{8-p}$ (except for the $D5$-brane, which yields again ${\cal M}_7
\xx S^3$).  As before, these metrics can not only be realised
asymptotically, but everywhere in spacetime after making the shift
transformation.
With the exception of $adS_5\times S^5$ these solutions
break $1/2$ of supersymmetry.
We note that the ${\cal M}_7\times S^3$ solution of type I supergravity
corresponding to the fivebrane with shifted harmonic function
must correspond to the $1/2$ supersymmetric ${\cal M}_7$ solution with linear
dilaton of gauged $N=2$ $d=7$ supergravity
\cite{SalSez} found in \cite{LPS}.
Also, the solution of a string in a fivebrane gives rise, after the shift,
to $adS_3\times E^4\times S^3$ geometry, and $adS_3\times
T^4$ is a solution of $N{=}2$ $d{=}7$ gauged supergravity as
well\cite{HKL}.
These observations strongly suggest that $N{=}1$ $d{=}10$ supergravity
compactified on a three-sphere yields 
$N{=}2$ $d{=}7$ gauged supergravity (see also \cite{DTN}).
Similar studies might be made for the other cases.

We have seen that dualities can connect spacetimes with different asymptotic
geometries, leading to several interesting applications.
A more detailed account of the issues reported here, and further
extensions, will be discussed elsewhere\cite{BPS}.

\section*{Acknowledgements}
We would like to thank I. Bakas, J. de Boer
and E. Kiritsis for useful discussions. KS is supported 
by the European Commision HCM program CHBG-CT94-0734.

\end{document}